# Temperature Runaway in a Pulsed Dielectric Barrier Discharge Reactor

H.Sadat[1], N.Dubus[1], J.M.Tatibouët[2]

[1]. Institut PPRIME, UPR CNRS 3346
[2]. Laboratoire de Catalyse en Chimie Organique, UMR CNRS 6503

Université de Poitiers, Ecole Supérieure d'Ingénieurs de Poitiers,
40, Avenue du Recteur Pineau, 86022 Poitiers (France)

**Abstract :**

This paper reports on experimental measurements of the gas temperature in a dielectric barrier discharge reactor powered by a high voltage pulsed signal. It is shown that the thermal behavior of the reactor follows a first order model. However, an unexpected runaway phenomenon was observed at a frequency of 300Hz. A sudden increase in the power source and consequently in reactor temperature which reaches 170°C is observed. This behavior is discussed in terms of input power variation with temperature, possibly due to a resonance phenomenon.

## 1. Introduction

Non thermal plasmas which can be generated at atmospheric or lower pressures are in non-equilibrium state so that the gas temperature is considered to be at a temperature only slightly higher than the ambient temperature while electrons temperature is about 10000K. In these plasmas, free radicals (O°, HO°...) are formed and can be used in many applications such as pollution abatement [1-3], surface treatment [4] or chemical vapor deposition (CVD) [5]. A great deal of experimental, theoretical and numerical works has been conducted so far and helped us to better understand these processes [6]. Whatever the concerned application, gas temperature has a great influence on the electrical parameters and on the nature and production of primary species formed by plasma and must then play an important role.

The increase of temperature leads to modify current and voltage shapes as it has been shown in [7]. Moreover, the increased number of gas collision can cause a large fraction of the input power to be dissipated in gas heating. The measurements have been generally performed by spectroscopic methods which allow the determination of vibrational and rotational molecules temperatures. The direct use of thermocouples in the plasma region is not possible since the electromagnetic environment and the presence of electrons and ionized species can initiate undesirable sparks. In a previous work we have already shown that gas temperature measurements can be achieved by using optical fibers [8]. These very first results have allowed us to calculate the heat transfers and losses with a simple analytical heat conduction model. A lumped heat capacitance model has been presented in [9] for the modeling of the unsteady regime.

According to the work of Okazaki et al. [10] and Liu et al. [11] the use of a pulsed power source seems to be more efficient for volatile organic compound treatment or ozone formation than sinusoidal one. We have therefore used a bipolar pulsed power supply which could produce positive and negative pulsed discharge so that the charges on the dielectric surface accumulated during the first part of the pulse discharge strengthen the electric field induced by the inversion of the polarity, favoring then the plasma formation.

In this work we report for the first time an experimental measurement of the gas temperature in a dielectric barrier discharge (DBD) plasma reactor powered by a bipolar voltage pulse with a frequency ranging from 100 to 300Hz which corresponds to the range of input energy densities used for VOC elimination [12, 13]. In the case of VOC elimination from air many works are published coupling a catalyst to the plasma reactor [13, 14], the efficiency of the catalyst increasing when it was heated. In order to reduce the energy expense or to avoid the thermal destruction of the reactor it is necessary to describe how the plasma alone can heat the reactor.

The unsteady electrode and gas temperatures were monitored from the initial time until a steady state was reached. It is shown that the temperature evolution is generally that of a first order model. The steady-state temperature increases with the input power from 49°C for P=0.9W (frequency=100Hz) to 56°C for P=1.8W (frequency=200Hz). When the frequency is increased to 300Hz, a temperature runaway effect is observed and the temperature can reach a value as high as 180°C in some situations.

## 2. Experimental setup

The experimental system mainly consists of a bipolar pulsed power supply, a coaxial DBD reactor and current and voltage measurement sets. The DBD reactor has been described elsewhere [11-12]. It presents a usual coaxial geometry. The reactor consists in a pyrex tube (12.4 mm in-diameter, 1.8 mm thickness) as dielectric barrier, an internal electrode (3.2 mm diameter steel tube) and an external electrode which is a 1mm thick and 80mm length copper foil wrapped around the pyrex tube. The volume in which occurs the plasma discharge (Fig.1) is equal to 9 cm$^3$. Plasma was generated by a bipolar pulsed electric generator (A2E Technologies). The voltage and current measurements were monitored with a digital oscilloscope (Lecroy CT3744, 500MHz) through high voltage probes (Lecroy, PPE 20kV, 100MHz) and an inductive current probe (Stangenes Industrie, 60MHz). The discharge power was calculated by time integration on a single period of the discharge voltage difference between the two electrodes and the measured current. Typical voltage and current profiles are shown on Figure 2.

Experiments were carried out at atmospheric pressure in a room where the temperature is fixed to 20°C. The flow rate of air (synthetic dry air from Air Liquide) through the reactor was fixed to 77 ± 1 mL/min (Brooks mass flow controllers). When the pulsed high voltage is applied, temperature begins to rise due to joule effect heating. The input power is balanced by various thermal losses so that the reactor reaches a thermal equilibrium. Temperature measurements inside the plasma zone and on the central electrode have been achieved by means of optical fibers connected to a Gallium Arsenide specific sensor. The temperature measurements were recorded at a frequency of 0.5 Hz by means of a specific trade sensor composed of a Y-bifurcated fiber provided with a gallium arsenide (GaAs) chip. Beyond 850 nm, gallium arsenide becomes optically transparent. The bandwidth of that component depends on the temperature and varies by an amount of 0.4nm/K. The position of the bandwidth, and therefore the temperature value, is determined by the spectral detector integrated to the measurement box. Temperature of the outer electrode was measured by an Infra-Red pyrometer thermometer (Raytech, MiniTemp MT4). Temperatures were recorded at different locations in the reactor with an accuracy of 0.1°C.

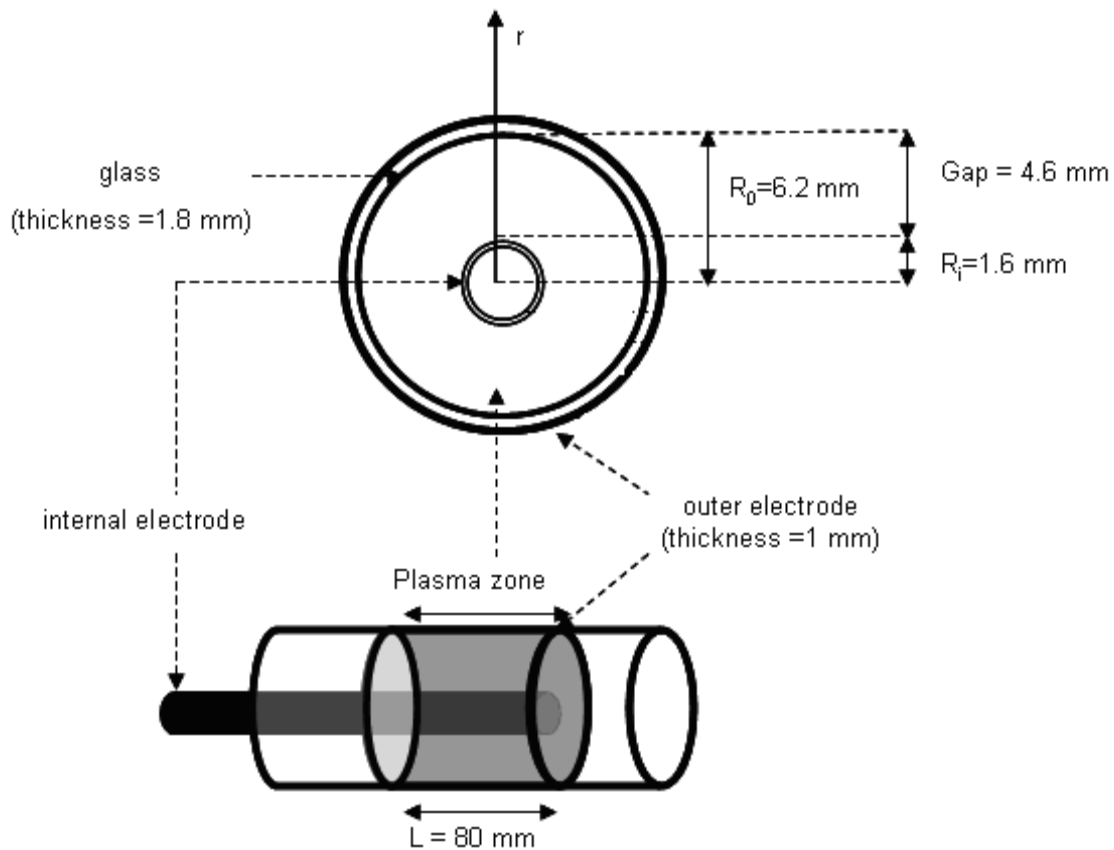

Figure 1: Scheme of the DBD reactor

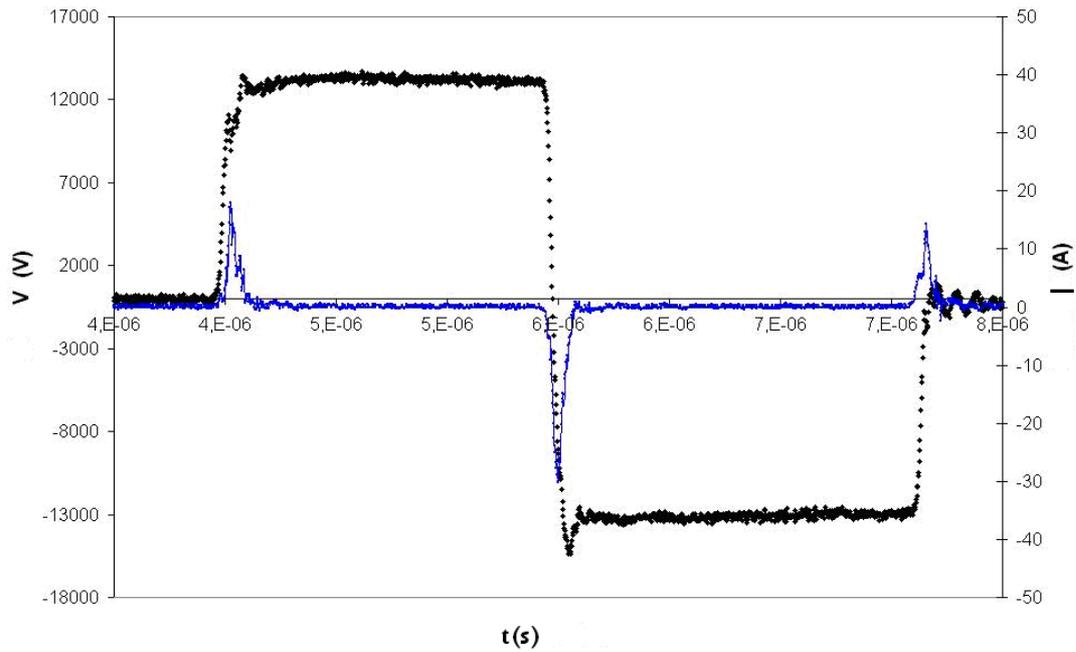

Figure 2: Voltage and current waveforms across the discharge

The input energy was determined by integrating the instantaneous power expression calculated by $P(t)=I(t).U(t)$ in function of time: $E=\int P(t).dt$ for the pulse duration.

## 3. Results

We present in this section the temperature evolution with time for three different frequencies, namely 100Hz, 200Hz and 300Hz.

Figure 3 shows the inner electrode temperature Te for a frequency of 100 Hz and an applied voltage of 13.4 KV. The temperature rises gradually with time and reaches a steady state value of nearly $Te_{lim}=40°C$ with a time constant of approximately 400 seconds. The gas temperature shows the same trend and reaches a steady state value of 30°C.

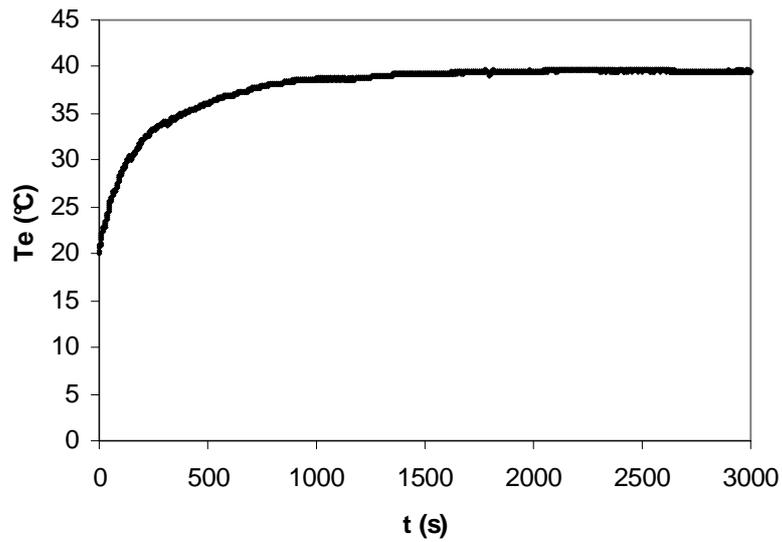

Figure 3: Electrode temperature versus time (100 Hz, 13.4 KV)

When the frequency is increased to 200Hz, the same trend is observed but the steady state value is higher (58°C for the electrode and 40°C for the gas) as shown on Figure 4 which presents the inner electrode temperature evolution.

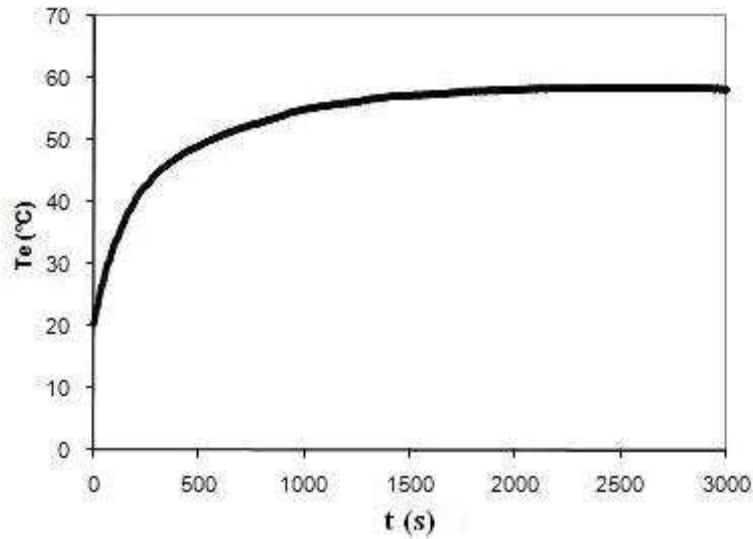

Figure 4: Electrode temperature versus time (200 Hz, 13.4 KV)

The electrode and gas temperatures are described by a first order model similar to the one described in [12] for the sinusoidal input power. As a matter of example, we present on figure 5 the function $\ln[(Te-Ts)/(Ti-Ts)]$ plotted versus the time (with Te=electrode temperature,

Ts=electrode temperature at steady state, Ti= initial temperature). The slope of the obtained straight line is equal to 0.0024, which corresponds to a time constant of 417 seconds.

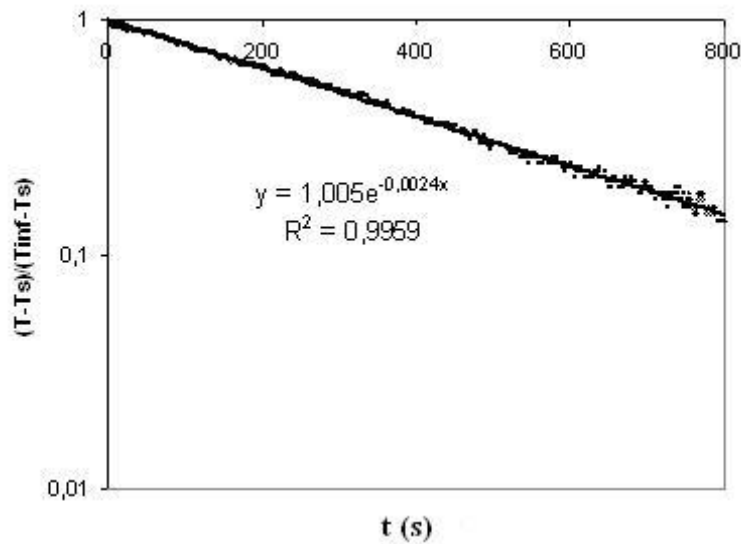

Figure 5: slope of the first order model

A further increase of the frequency up to 300Hz leads however to an unexpected phenomenon. At this frequency we have observed that the temperature of the inner electrode begins to rise gradually with the same behavior as previously observed at 100 and 200 Hz, but a fast increase in the temperature appears suddenly when the electrode temperature reaches a value of about 60°C. The profile of the temperature evolution of the inner electrode as a function of time is presented figure 6. This behavior clearly indicates that two steps can be distinguished in the temperature evolution. The first one which starts at ambient temperature until about 65°C, follows approximately the same behavior than the temperature profiles observed for the 100 and 200Hz pulse repetition frequencies.

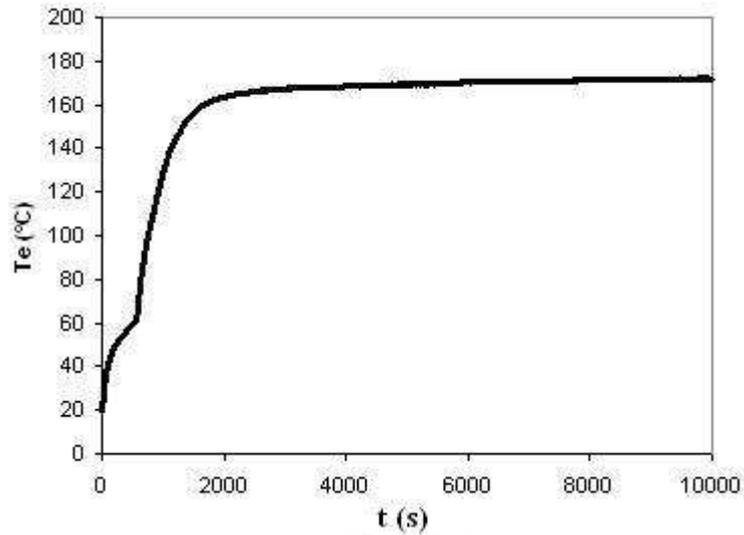
Figure 6: Electrode temperature for 300Hz and 13.4KV

However, as shown on figures 6 and 7, an inflexion of the curve occurs suddenly when the temperature reaches 60°C and a faster growth of temperature takes place (although the general evolution is still that of a first order model).

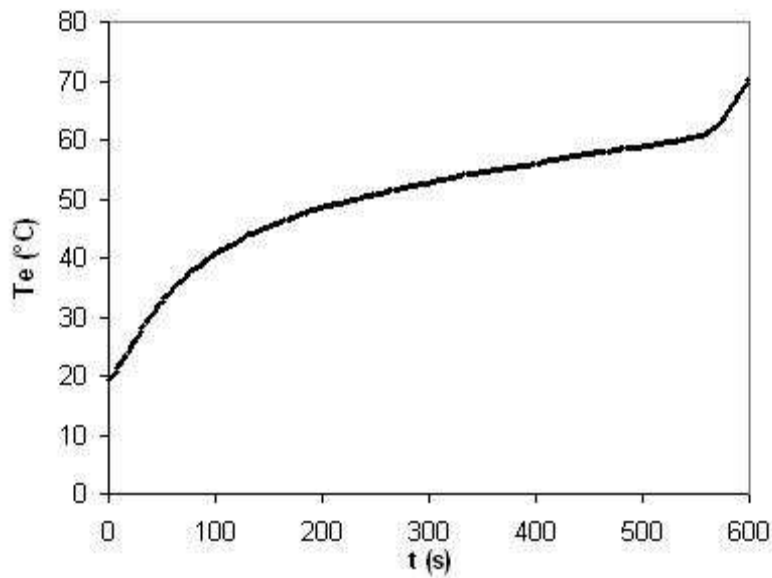
Figure 7: Electrode temperature: early stage

A gradual variation of the electrode temperature is then observed until a steady state of 170°C is reached. The same trend is observed for the gas temperature variations. This clearly shows that the power used to heat the gas increases suddenly.

## 4. Discussion

When an electrical power is delivered to a DBD reactor, a fraction of the energy contributes to heat the gas and the reactor whereas the other part is used by the chemical transformations of the gas molecules (ionization, chemical bonds breaking, activation...). According to previous experiments, the only detectable products formed by plasma in dry air are ozone and nitrogen oxides at a concentration which does not exceed 1000 ppm. A rough calculation based on these data and on our experimental conditions allows estimating the energy needed to the chemical processes by using the following reactions:

$3/2 \, O_2 \rightarrow O_3 \quad \Delta H = 142.7 \text{ kJ/mol} \quad \text{and} \quad 1/2 N_2 + 1/2 O_2 \rightarrow NO \quad \Delta H = 91.3 \text{ kJ/mol}$

We found 0.008 W and 0.005W for ozone and NO formation, respectively. These values mean that only a very small part of the input energy is used for the chemical processes.

Due to the low thermal isolation of our reactor, a large part of the heat is dissipated to the outdoor of the reactor by conduction, radiative transfer and by heating the gas passing through the reactor, so that a thermal transfer equilibrium is reached, leading to stable temperatures of the gas and of the various parts of the reactor after a stabilization time.

An unexpected phenomenon is observed when the temperature exceeds a threshold value (65°C in the particular case of this study) which is only obtained in our conditions when the pulse repetition frequency is 300Hz. In these conditions, the temperature of the inner electrode and of the gas increases with time following the same kind of profile than for the other frequencies (100 and 200Hz) but, as soon as the threshold temperature is reached (≈ 65°C) a fast increase of the temperature is observed until a plateau was reached at a value as high as 175°C.

Another series of experiments have been carried out by using a modified reactor where the active region has been surrounded by an external glass envelope. Furthermore, we have deposited a highly reflecting metallic film on the internal surface of the envelope in order to diminish the radiative thermal heat transfer to the environment. The electrode temperature of the insulated reactor (with 300Hz and 11.9kV) is presented on figure 8. The same phenomenon is observed. The inflexion of the curve is near 53°C and the steady state temperature is 185°C.

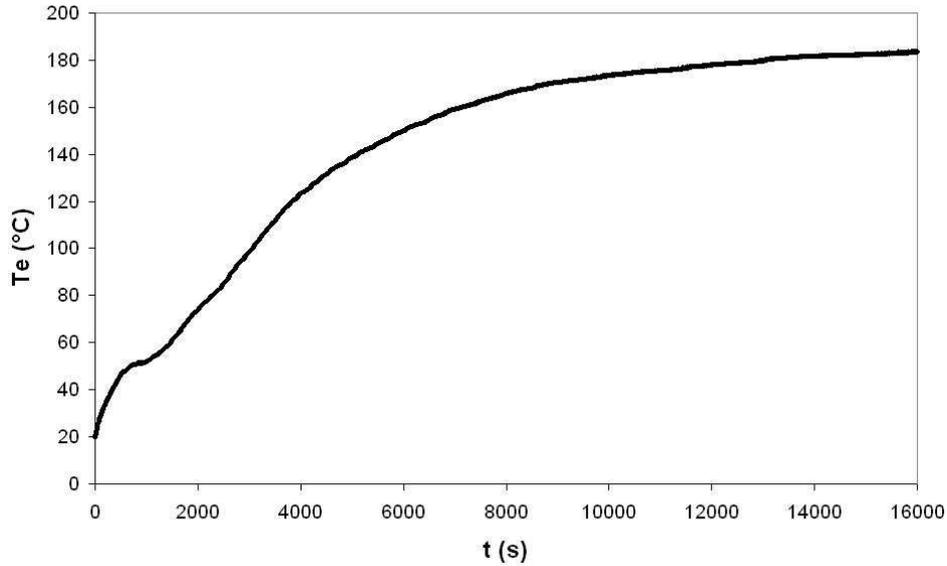

Figure 8: Electrode temperature for the insulated reactor (11.9 KV 300 Hz)

The same thermal behavior has been observed with this reactor, but with a higher value of the steady-state temperature due to the better insulating of the reactor. This behavior could be partly explained by an increase in the input energy due to the increase of the gas temperature as shown in the figure 9.

The fast increase in the temperature could indicate a change in the nature of the plasma from homogeneous-like with a small number of streamers to highly filamentous plasma with a high number of streamers. The presence of a high number of streamers enhances the gas ionization by plasma leading to a decrease in its electrical resistivity favoring then a high value of the plasma current and by consequence an increase in the plasma deposited energy in the gas.

The measured input power shown on figure 9 increases with time and reaches the steady state value of P=10.5W.

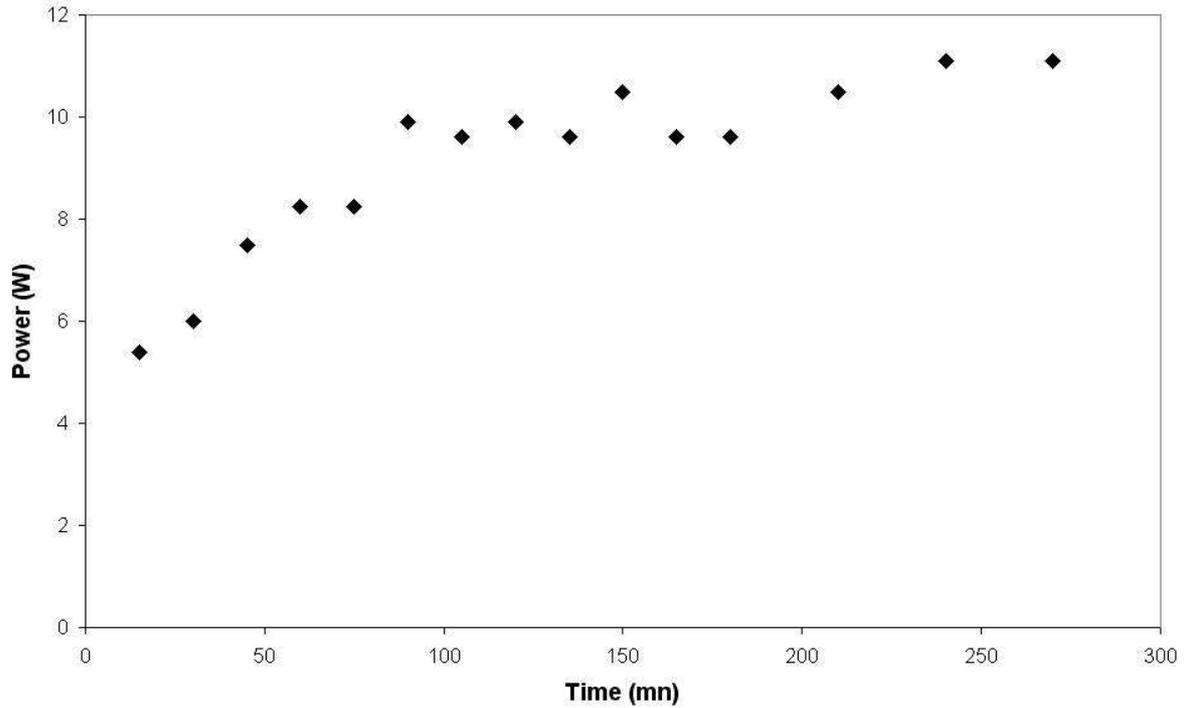

Figure 9: Power versus time for the insulated reactor

It remains to explain why the input power increases in such a way with temperature. First of all, it must be noted that the increase of the gas temperature leads to a decrease of the gas density and therefore to an increase of the reduced electric field. More reactive streamers can then be formed and the required maximum peak ignition voltage is also reduced. Next, by an increase of the temperature, the reactor impedance decreases and more power is used for streamer generation.

Moreover, we can consider the plasma reactor as a RLC circuit whose impedance varies with temperature so that the system can move to the electrical resonance domain, as shown by Garamoon et al. for a similar plasma system [15]. In the resonance domain conditions, these authors have observed that the input power increases quickly and pass through a maximum corresponding to the resonance conditions. If we suppose that our system is close to the resonance domain conditions at a frequency of 300 Hz, the slight variation of impedance due to the increase of temperature could be enough to reach the resonance conditions and then to rapidly increase the power input leading then to observe the fast temperature increase shown in figure 9. Let us mention finally that the authors in [16] have also reported an increase in the discharge power with temperature in their reactor.

## 5. Conclusion

The measurement of the inner temperature of a DBD gas discharge reactor has shown that it increases when the plasma is on until a steady state was reached. The thermal behavior of the reactor follows a first order model with a time constant close to 400 seconds.

The steady-state temperature increases with the input power from 49°C for P=0.9W (frequency=100Hz) to 56°C for P=1.8W (frequency=200Hz).

An unexpected phenomena was observed when the frequency of the pulse repetition was 300 Hz since after an increase in temperature as a function of time similar to those recorded for 100 and 200 Hz, a sudden increase both in the input power and in the temperature was observed when the temperature attained a threshold level depending of the thermal insulation of the reactor (65°C for the basal reactor and 53°C for the thermally insulated one). This unexpected phenomenon could be explained by the possible modification of the impedance of our plasma reactor as a function of temperature, leading to be in the conditions of an RLC circuit resonance phenomena.